\documentclass[prb,superscriptaddress,twocolumn,showpacs,amsmath,amssymb]{revtex4}% Physical Review B

\usepackage{graphicx}% Include figure files
\usepackage{dcolumn}% Align table columns on decimal point
\usepackage{bm}% bold math

\begin{document}

\title{Calculation of overdamped c-axis charge dynamics and the coupling to polar
phonons in cuprate superconductors.}
\author{W. Meevasana}
\email[]{non@stanford.edu}
 \affiliation {Department of Physics,
Applied Physics, and Stanford Synchrotron Radiation Laboratory,
Stanford University, Stanford, CA 94305}

\author{T.P. Devereaux}
\affiliation {Department of Physics, University of
Waterloo,Waterloo, Ontario, Canada N2L 3G1}

\author{N. Nagaosa}
\affiliation {CREST, Department of Applied Physics, University of
Tokyo, Bunkyo-ku, Tokyo 113, Japan}

\author{Z.-X. Shen}
\affiliation {Department of Physics, Applied Physics, and Stanford
Synchrotron Radiation Laboratory, Stanford University, Stanford,
CA 94305}

\author{J. Zaanen}
\altaffiliation{On leave of absence from the Instituut-Lorentz for
Theorectical Physics, Leiden University, Leiden, The Netherlands}
\affiliation {Department of Physics, Applied Physics, and Stanford
Synchrotron Radiation Laboratory, Stanford University, Stanford,
CA 94305}

\date{October 4, 2006}% It is always \today, today,

\begin{abstract}
In our recent paper we presented empirical evidences suggesting
that electrons in cuprate superconductors are strongly coupled to
unscreened c-axis polar phonons. In the overdoped regime the
c-axis metallizes and we present here simple theoretical arguments
demonstrating that the observed effect of the metallic c-axis
screening on the polar electron-phonon coupling is consistent with
a strongly overdamped c-axis charge dynamics in the optimally
doped system, becoming less dissipative in the overdoped regime.

\end{abstract}

\pacs{71.38.-k, 74.72.Hs, 79.60.-i}% PACS, the Physics and Astronomy
\maketitle

\section{\label{introp}Introduction}

As electrodynamical media the cuprate high-T$_c$ superconductors
are highly anomalous. In their normal state they behave like
metals in the planar ab-directions while they can be regarded as
dielectrics along the interplanar c-direction, at least at finite
frequencies. Only in the regime where the superconductivity start
to degrade because of too high doping levels, metallic `normalcy'
develops along the c-axis \cite{gencaxis}.

Electrodynamical 'anomaly' raises a general question regarding the
nature of electron-phonon coupling in the cuprates. Insulating
cuprates are as any other oxide characterized by a highly
polarizable lattice. The consequence is that one is dealing with
strongly `polar' electron-phonon (EP) couplings, associated with
the long range nature of the Coulomb interaction between carriers
and the ionic lattice \cite{MullerBednorz}. In normal metals these
long range interactions are diminished by metallic screening and
one is left with rather weak, residual short range EP
interactions. Although this is also the case in the planar
directions of the cuprates, due to incoherent electron motion
perpendicular to the planes we argue that the c-axis phonons are
largely unscreened in this direction and strong polar EP-couplings
are active \cite{Falter}.

In our recent paper \cite{Non:experiment}, angle-resolved
photoemission (ARPES) studies of optimally-doped and overdoped
Bi$_{2}$Sr$_{2}$CuO$_{6}$ (Bi2201) superconductors have indicated
that polar c-axis couplings play an important role. Energy
dispersion "kinks" resulting from coupling to well defined bosonic
modes, which we argued to be phonons rather than spin resonance
modes, are weaker at low binding energies in overdoped compared to
optimally doped Bi2201. This case is twofold: (a) a good
description is obtained for the electronic self-energy in the
nodal momentum direction in an optimally doped single layer Bi2201
superconductor by assuming that the Eliashberg function $\alpha^2
F(\omega)$ largely coincides with the measured c-axis electron
energy loss function; (b) The electronic self-energy drastically
changes in a strongly overdoped system and we show that these
changes are accounted for assuming a frequency dependent screening
characterized by a scale $\omega_{scr,c} \simeq 60$meV. Here we
will present calculations further substantiating this claim. We
will give the reasons why the c-axis optical loss function should
be a good model for the Eliashberg function associated with the
polar couplings in the optimally doped system. Furthermore, we
will show that the way the self-energy diminishes in the overdoped
regime is qualitatively- and quantitatively consistent with what
has been learned from optical measurements about the way
metallicity develops in the overdoped regime.

We remark that a great deal of focus has been placed on in-plane
phonons such as the breathing modes showing large softenings with
doping in the cuprates \cite{breathing}. These modes have been
shown to contribute to the electronic self-energy for nodal
electrons \cite{Zhou}. Moreover, the breathing phonons provide
signatures on the spectral function for the formation of charge
ordering \cite{charge_ordering} and may contribute to $d-$wave
pairing in the presence of strong correlations \cite{ishihara}.
Rather than taking into account the interesting and relevant issue
of planar phonon couplings, our focus here will be on the $c-$axis
phonons in order to explore the role of screening.

The c-axis is insulating, not because a lack of carriers, but
instead due to the fact that charge dynamics is strongly
overdamped and incoherent. To be discussed at the end, the damping
rate is decreasing rapidly in the overdoped regime, and this is
exactly what is needed to explain the behavior of the self-energy.
Van der Marel and Kim \cite{PlasmonInOut:Marel} pointed out some
time ago that the \emph{effective} c-axis plasmon frequency
$\omega_{p,c} = 4 \pi n_c e^2 /m_c$ ($n_c$ and $m_c$ are the
c-axis effective carrier density and band mass, respectively) as
determined by an optical sum rule can be as large as 1/4 of the
planar plasma frequency $\omega_{p,ab} \simeq 1$ eV
\cite{InplanePlasma:Marel}. The essence of overdamped dynamics is
captured by the expression for the dielectric function associated
with a plasmon with damping rate $\Gamma$ and frequency
$\omega_{p,c}$ at long wavelengths,
\begin{equation}
\epsilon_{el} ( \omega) = \epsilon_{\infty} - \frac {
\omega_{p,c}^2} {(\omega^2 + i \omega \Gamma)} \label{plasmoneps}
\end{equation}
It is demonstrated (see Eq. (\ref{charfreq2}) that in the
overdamped case $\Gamma >> \omega_{p,c}$ the medium behaves like a
dielectric at frequencies larger than the characteristic
frequency,
\begin{equation}
\omega_{scr,c} \approx \frac { \omega_{p,c}} {\epsilon_{\infty}
\Gamma} \; \omega_{p,c} \label{charfreq}
\end{equation}

In the cuprates, it will be shown that in the doping range up to
optimal doping, this frequency is smaller than the typical optical
phonon frequencies with the effect that the c-axis optical
conductivity $\sigma^1_c (\omega)$ looks like that of a polar
insulator, dominated by unscreened polar phonons. To get the
impression of the magnitude of $\omega_{scr,c}$, we assume a
simple Drude form for the optical conductivity (see Eq.
(\ref{optcon})). With this form, $\omega_{scr,c}$ is approximately
proportional to the height of the electronic background in the
c-axis $\sigma^1_c (\omega)$ and this background is found to
increase rapidly in the overdoped regime in LSCO system
\cite{LSCO:Tajima}. We find that this change is just of the right
magnitude to explain the changes in the self-energy of the
overdoped system.

The justification for taking the \emph{c-axis} EP couplings being
representative of the interactions showing up in the self-energy
will be discussed first. The actual importance of polar
EP-couplings in the cuprates is \emph{a priori} a rather subtle
affair. On the one hand, there are actually very few phonons which
can be regarded as unscreened. The cuprates can be viewed as a
stack of metallic sheets. For one limit, when considering phonons
with $\vec{q}$ purely along c-axis ($\vec{q}=\vec{q}_{c}$), the
screening frequency $\omega_{scr,c}$ should be set by the
effective c-axis plasmon frequency $\omega_{p,c}$, such that all
phonons with frequency $\omega_{ph, i} (\vec{q}=\vec{q}_{c}) <
\omega_{scr,c}$ can be regarded as screened. This $\omega_{scr,c}$
will be later shown to be defined by Eq. (\ref{charfreq2}).
Similarly, for the other limit, when considering phonons with
$\vec{q}$ purely in the ab-plane ($\vec{q}=\vec{q}_{ab}$),
$\omega_{scr,ab}$ should be set by the planar plasmon frequency
$\omega_{p,ab}$. Since $\omega_{p,ab}$ is large compared to
$\omega_{p,c}$ \cite{InplanePlasma:Marel}, this implies that
phonons with $\vec{q}$ along the c-axis will be left unscreened
compared to phonons with $\vec{q}$ in ab plane. In a more general
case, at a 3D momentum $\vec{q} = \vec{q}_{ab} + \vec{q}_c$, the
effective 3D plasmon frequency $\omega_{p,\vec{q}}$ for $\vec{q}
\rightarrow 0$ \cite{Falikov,PlasmonInOut:Marel} (also see
Appendix A) is given by,
\begin{equation}
\omega^2_{p,\vec{q}} =  \frac{ q_c^2} {q^2} \omega_{p,c}^2 +
\frac{ q_{ab}^2} {q^2} \omega_{p,ab}^2 \label{3DOmega}
\end{equation}
The polar coupling requires that the phonon frequencies exceed the
3D screening frequency, which should be set by this
$\omega^2_{p,\vec{q}}$. Hence, this implies that only phonons in a
narrow cone around $q_{ab} \simeq 0$ can contribute to the polar
couplings. It is questionable whether this small $q_{ab}$ will
show up in the polar couplings. Given the unknowns, this question
is not easy to answer on theoretical grounds. On the other hand,
leaving the overall strength aside, the frequency dependence of
the electronic self-energy can be easily calculated
\cite{Mahan,eliashberg:Grimvall} and in section III we will
compare the experiment \cite{Non:experiment} with the calculation,
assuming the predominant electron-phonon coupling is due to the
unscreened c-axis phonons.

\section{\label{resigmap}REAL PART OF ELECTRON SELF-ENERGY}

In the following, the real part of self-energy, $Re(\Sigma)$,
contributed from interactions which can be captured by the c-axis
optical dielectric function, $\epsilon_{c}(\omega)$, will be
calculated where the strength of interactions as a function of
energy is represented by the Eliashberg function, $\alpha^2
F(\omega)$. Regarding the empirical result shown in the
experimental paper \cite{Non:experiment}, the main results in this
paper are that a) the c-axis loss function, $Im (-1/\epsilon_c
(\omega))$, is a good representative of $\alpha^2 F(\omega)$ for
the optimally doped cuprate and b) when including screening
effect, the expression of $\alpha^2 F(\omega)$ becomes Eq.
(\ref{elias2metph}) giving the good agreement of $Re(\Sigma)$ of
the overdoped cuprate.

The electron self-energy can be given by ($\delta \rightarrow
0^+$),
\begin{eqnarray}
&&\Sigma (\vec{p}, \epsilon) = - \frac {1} { (2 \pi)^4 \pi}\int
d^3p_1\int d\omega\int d\epsilon_1 Im W(\vec{p} - \vec{p}_1,
\omega) \nonumber \\ &&\times\frac {Im G ( \vec{p}_1, \epsilon_1)}
{\omega+ \epsilon_1 - \epsilon - i \delta} \left(
\tanh\frac{\epsilon_1}{ 2k_B T} + \coth\frac{\omega} { 2k_B T}
\right) \label{GWself}
\end{eqnarray}
where $G$ and $W$ are the retarded electron propagator and the
effective interaction, respectively. Here we neglect the detailed
momentum structure of the electron-phonon coupling, which although
important for a subset of phonons (such as the B$_{1g}$ and
B$_{1u}$ c-axis phonons), is motivated by the interest in small q
phonons providing a more or less isotropic charge displacement
along the c-axis. We neglect vertex corrections which may be
important for polaron formation but are not crucial to develop the
ideas of screening.

When only considering the interaction of electrons to polar c-axis
phonons and electrons to electrons, the effective interaction (see
Appendix B) can be written as
\begin{equation}
W_{eff} (\vec{q}, \omega ) = \frac { v_q } { \epsilon_{tot}
(\vec{q}, \omega)  } \label{Wtoeps}
\end{equation}
where $\epsilon_{tot} (\vec{q}, \omega)=\epsilon_{\infty} - v_q
P_e ( \vec{q}, \omega ) + \epsilon_{ph} ( \vec{q}, \omega )$ is
the total dielectric function in which the core polarization, the
phononic- and electronic polarizations, in terms of the electronic
polarizability $P_e ( \vec{q}, \omega)$, enter additively.

The effective interaction includes both Coulomb and EP
interactions, and the manifestation of screening is that the
interactions are mixed as Coulomb interactions modify the phonon
propagator $D$ and vice-versa. This effective interaction can be
rewritten as \cite{Mahan} ($M_q$ is the bare EP vertex neglecting
any fermionic momentum dependence)
\begin{equation}
W_{eff} (\vec{q}, \omega) = \frac{ v_q} {\epsilon_{el} ( q,
\omega)} + \frac{ \epsilon_{\infty}^2 M^2_q  D (q, \omega)}
{\epsilon_{el}^2 (q, \omega)} \label{effint}
\end{equation}
where $\epsilon_{el}$ is the electronic polarizability Eq.
(\ref{plasmoneps}) and $v_q=4\pi e^{2}/q^{2}$; this is a general
expression for phonon-modulated uniform charge displacements.

We now assert that the effective interaction $W_{eff}$ is
dominated by the coupling to the unscreened c-axis phonons. Next
we assume that momentum dependences are smooth. This is definitely
not problematic for the phonon-propagators hidden in $W_{eff}$
because the important phonons are of the high energy optical
variety having small dispersions. This is further helped by the
fact that because of the metallic screening in the planar
directions, only phonons with a small planar momentum $q_{ab}$ can
be regarded as unscreened. A more delicate issue relates to the
electron momentum dependence of the bare vertex having
implications for how the self-energy will depend on $\vec{p}$.
Elsewhere we will show that this can give rise to highly
interesting effects associated with the screening physics
\cite{smallq:Tom} but here we will focus on the frequency
dependences in a given momentum direction.

Neglecting momentum dependences, the phonon polarizability can be
straightforwardly parameterized, by a sum over phonons with
transversal frequencies $\omega_{Tj}$, damping $\gamma_j$ and
strengths $s_j$, as tabulated by Tsvetkov {\em et al.}
\cite{LossFunction:Marel},
\begin{equation} \epsilon_{ph} (\omega)
= \sum_j  \frac{ s_j \omega^2_{Tj}} { \omega^2_{Tj} - \omega^2 + i
\omega \gamma_j} \label{phoneps}
\end{equation}

Taking for the electronic polarizability $v_q P_e (\omega) =
\omega^2_{p,c} / \omega^2$ and a plasmon frequency $\omega_{p,c}$
large compared to all phonon frequencies one recovers the standard
results for the EP coupling in metals, but we now focus on the
category of unscreened phonons, such that $v_q P_e (\omega) = 0$.

In the insulator, the phonon propagator turns into the bare phonon
propagator $D \rightarrow D^0$ which becomes in terms of the
parametrization Eq. (\ref{phoneps})
\begin{equation}
M^2_{q \simeq 0} D^0 (\omega) =-v_{q \simeq 0}^\infty
\frac{\epsilon_{ph}(\omega)}{\epsilon_\infty +
\epsilon_{ph}(\omega)} \label{bphonprop},
\end{equation}
where $v_{q}^\infty=v_{q}/\epsilon_\infty$.

The Eliashberg function is defined as the Fermi-surface average
\cite{Mahan},
\begin{equation}
\alpha^2 F_{\vec{k}} (\omega) = \frac { 1} { (2\pi)^3} \int \frac{d^2 k'}
{v_F} |  \frac{ \epsilon_{\infty} M_{\vec{k}-\vec{k'}}}
{\epsilon_{el} (\vec{k}-\vec{k}', \omega)}  |^2 Im D (\vec{k} -\vec{k'}, \omega)
\label{eliasdef}
\end{equation}
while the real part of the electronic self-energy, calculated from
the Eliashberg function, is given by \cite{eliashberg:Grimvall},
\begin{equation}
Re \Sigma (\vec{p}, \epsilon) = \int_0^\infty d\omega
\alpha^2F(\omega; \vec{p},
\epsilon)K\left(\frac{\epsilon}{kT},\frac{\omega}{kT}\right)
\label{Eliashberg}
\end{equation}
where $K(y,y')=-\int_{-\infty}^\infty dx f(x-y)2y'/(x^2-y^2)$ with
$f(x)$ being the Fermi distribution function.

\subsection{\label{unscreened}$\alpha^2 F (\omega)$ for c-axis insulating system (optimally-doped)}

Assuming smooth momentum dependences, and observing that in the
insulator $\epsilon_{el} = \epsilon_{\infty}$, one immediately
infers that the Eliashberg function for the polar phonons can be
written as,
\begin{equation}
\alpha^2 F(\omega) = b \frac{|M^2_{q \simeq 0}|}{v_{q \simeq 0}}
Im D^0 (\omega) = b\cdot Im \left( - \frac {1} {\epsilon_{tot}
(\omega)} \right) \label{elias2eps}
\end{equation}
where lumping together in the factor $b$ is the numerical factors
coming from the momentum integrations (see Appendix C). Since
momentum dependences are weak we can as well take the optically
measured c-axis loss function $Im (-1/\epsilon_c (\omega))$ where
$\epsilon_c (\omega) = \epsilon_{tot} (q_{ab} = 0, q_c \rightarrow
0, \omega)$ as representative for the loss functions at small
$q_{ab}$ and arbitrary values of $q_c$. In this way, we obtained a
good fit to the measured self-energy in the optimally doped
cuprates \cite{Non:experiment}.

\subsection{\label{screened}$\alpha^2 F (\omega)$ with metallic screening
effect (overdoped)}

In the following, we will include the metallic screening effect to
calculate $Re(\Sigma)$ in the overdoped case. In the experimental
paper \cite{Non:experiment} we found that the self-energy in the
overdoped system was reproduced in detail,
\begin{equation}
\alpha^2 F(\omega) =  b \frac {\omega^2} {\omega^2_{src,c} +
\omega^2}
 Im \left( - \frac {1} {\epsilon_c (\omega)} \right)\label{eli2fil}
\end{equation}
using the loss function of the optimally doped system and a
`filter' function representing a characteristic `screening'
frequency $\omega_{src,c} \simeq 60$meV.  As we will now argue,
the `filter' function $\omega^2 / (\omega^2_{src,c} + \omega^2)$
is consistent with the changes in the screening seen in optical
spectroscopy. To incorporate the effect of the metallic screening,
one could model the c-axis polarizability according to Eq.
(\ref{plasmoneps}) as,
\begin{equation}
v_{q_{//c}} P_e ( q_{//c}, \omega) = \frac { \omega_{p,c}^2}
{\omega^2 + i \omega \Gamma}
\end{equation}

We have neglected the details of the frequency and fermionic
momentum dependence of the scattering rate. This simple Drude form
is an over simplification but all what matters for the phonon
screening are the overall energy scales of screening and the
phonon frequencies. Moreover, we use "damping" as a way of loosely
parameterizing the change from coherent metallic transport along
the $c-$axis in overdoped systems to incoherent insulating
behavior near optimal doping. The nature of this change is a very
important issue, relating to charge confinement and/or the
incoherent nature of anti-nodal quasi-particles governing $c-$axis
propagation in an LDA treatment. A full description of this change
is difficult around optimal doping where strong correlations are
developed. An adequate description of the role of correlations
requires treating both electronic correlations and EP interactions
on equal footing. Since such treatments are presently limited, we
give only a heuristic picture of the consequences of strong
damping and incoherence.

For finite $\omega_{p,c}$, $1/\epsilon_{el}$ will also acquire an
imaginary part which contributes to the Eliashberg function. When
the plasmon is underdamped while $\omega_{p,c}$ is in the phonon
frequency window, the plasmon would become well `visible' in the
self-energy \cite{Classicplas}. However, in the overdamped regime,
$\Gamma \geq \omega_{p,c}$, of interest to the c-axis of the
cuprates, this turns into a smooth background contribution. In the
experimental determination of the self-energy such backgrounds are
subtracted. Although it would be interesting to find out if such
backgrounds associated with the c-axis metallization can be
extracted from the experimental data, these should be ignored in
the present context and we should focus on the phonon contribution
to the effective interaction in Eq. (\ref{effint}),
\begin{equation}
W_{sc-ph} ( \vec{q}, \omega) = \frac { M^2_q D ( \vec{q}, \omega
)} {\epsilon^2 (\vec{q}, \omega)} \label{phonint}
\end{equation}
where $\epsilon(\vec{q}, \omega) =
\epsilon_{el}/\epsilon_{\infty}$ and $\epsilon_{el}$ is modeled
according to Eq. (\ref{plasmoneps}). The renormalized phonon
propagator \cite{Mahan} is,
\begin{equation}
D (\vec{q}, \omega ) = \frac{ D^0 (\vec{q}, \omega ) } { 1 - M^2_q
D^0 (\vec{q}, \omega ) P_e (\vec{q}, \omega )/ \epsilon (\vec{q},
\omega)} \label{phonprop}
\end{equation}

By using Eq. (\ref{eliasdef}) and the same assumptions as for the
optimally doped case, the Eliashberg function including the
screening effect becomes,
\begin{equation}
\alpha^2 F (\omega) = b\frac{|M^2_{q \simeq 0}|}{v_{q \simeq 0}}
\; \frac{ 1}{|\epsilon(\omega)|^2} \; Im D (\omega)
\label{elias2metph}
\end{equation}

This more general form does reproduce the self-energies of both
the overdoped and optimally-doped systems quite well, as long as
the c-axis charge dynamics is not turning strongly underdamped.
\begin{figure}[t]
\includegraphics [width=3.2in, clip]{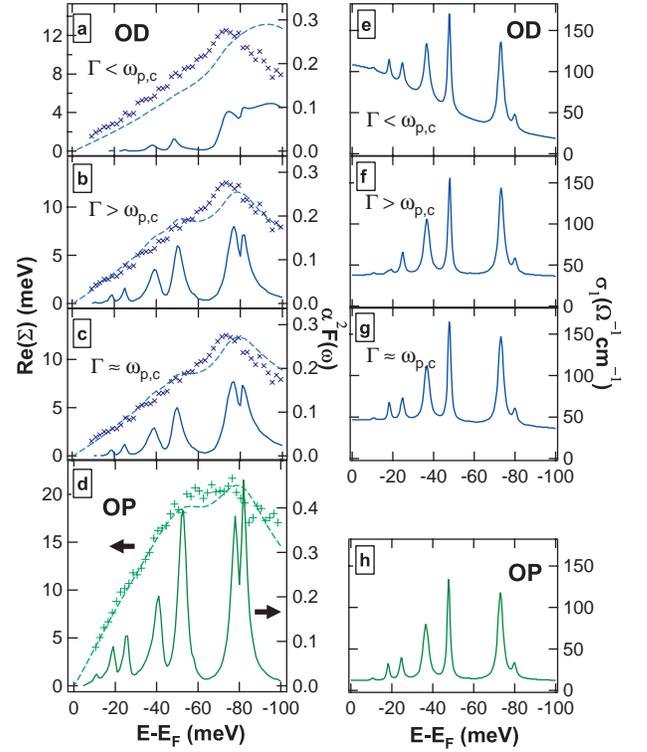}
\caption{\label{fig1:result} The experimental c-axis optical
conductivity in optimally doped (OP) from Ref. 16, (h). Various
models for the optical conductivity of overdoped (OD) Bi2201
(e)-(g). The corresponding Eliashberg functions (solid line) and
electronic self-energies (dash line) assuming that the predominant
electron-phonon coupling is due to the unscreened c-axis phonons
(a)-(d); experimental data in symbol are from Ref. 4. Parameters:
(a)-(c), $\omega_{scr,c} \approx 60$ meV. (a) $\Gamma
<\omega_{p,c}, \Gamma = 45$ meV and $\omega_{p,c} = 190$ meV, (b)
$\Gamma>\omega_{p,c}, \Gamma = 430$ meV and $\omega_{p,c} = 350$
meV, and (c) $\Gamma \approx \omega_{p,c}, \Gamma = 180$ meV and
$\omega_{p,c} = 250$ meV. (d) OP case, $\Gamma = 700$ meV and
$\omega_{p,c} = 250$ meV.}
\end{figure}

Given the  $\alpha^2 F (\omega)$ as in Eq. (\ref{elias2metph}),
the factor  $| \frac{ 1}{\epsilon (\omega)}|^2$ acts like a filter
function in the overdamped regime. The characteristic screening
frequency $\omega_{scr,c}$ can be determined by demanding $|
\frac{ 1}{\epsilon (\omega_{scr,c})}|^2 = 1/2$ such that the
phonons with frequencies less than $\omega_{scr}$ can be regarded
as screened. Since $\epsilon(\omega) = \epsilon_{el}
(\omega)/\epsilon_{\infty}$ it follows immediately from Eq.
(\ref{plasmoneps}),
\begin{equation}
\omega_{scr,c} = \frac{1}{\sqrt{2}}\left[\left(
\left(\Gamma^2+2\omega_{p,c}'^2\right)^2
+4\omega_{p,c}'^4\right)^{1/2}-\left(\Gamma^2+2\omega_{p,c}'^2\right)\right]^{1/2}\label{charfreq2}
\end{equation}
where $\omega_{p,c}'= \omega_{p,c}/\sqrt{\epsilon_{\infty}}$. In
the overdamped regime ($\Gamma >> \omega_{p,c}$),  $\omega_{scr,c}
\approx \frac { \omega_{p,c}} {\epsilon_{\infty} \Gamma} \;
\omega_{p,c}$ as in Eq. (\ref{charfreq}).

This $\omega_{scr,c}$ can be directly related to the electronic
background in the optical conductivity, $\sigma^1_c (\omega)$. Eq.
(\ref{plasmoneps}) leads to a simple Drude form for the
conductivity which can be expanded in the regime $\Gamma >>
\omega$ as,
\begin{equation}
\sigma_1(\omega) =
\frac{\omega_{p,c}^2}{4\pi\Gamma}(1-\frac{\omega^2}{\Gamma^2}+...)\label{optcon}
\end{equation}
and it follows from Eq. \ref{charfreq} that the height of this
background corresponds to $\omega_{p,c}^2/4\pi\Gamma = (
\epsilon_\infty / 4\pi) \omega_{scr,c}$ in the overdamped regime
($\Gamma >> \omega_{p,c}$). One notices that in principle the
self-energy could be determined using the information in the
optical conductivity, avoiding any free parameters. We note that
since the optical data on overdoped Bi2201 are not available and
we model these according to what has been measured in the
La$_{2-x}$Sr$_x$CuO$_4$ (LSCO) system.

\section{\label{exp} COMPARISON WITH EXPERIMENTAL DATA}

According to Ref. 9, the ab plasma frequency $\omega_{p,ab}
\approx 1$ eV while $\omega_{p,ab}/\omega_{p,c} \approx 4$ to $10$
and we take $\omega_{p,c}\approx\omega_{p,ab}/4\approx 250$ meV.
To reconstruct the measured optical conductivity in the OP case
\cite{LossFunction:Marel} [Fig. 1(h)] we need a large $\Gamma \sim
700$meV to obtain a background $\approx$ 12 $\Omega^{-1}cm^{-1}$.
In Fig. 1(f)-(g) we combine the (dressed)  phonon spectrum of
Bi2201 with the background  as measured in the LSCO system at 30\%
doping \cite{LSCO:Tajima}, translating in an
$\omega_{scr,c}\approx60$ meV, indicating the insensitivity to the
precise values of $\omega_{p,c}$ and $\Gamma$ separately. In Fig.
1(b)-(d) we show the outcomes for both the Eliashberg functions
and the calculated self-energies, finding results which closely
track the phenomenological filter function used in Ref. 4.

For completeness we include the outcomes assuming a strongly
underdamped plasmon [Fig. 1(a) and (e)]. One sees immediately that
this yields a much less satisfactory outcome. We also notice that
in the overdamped cases Fig. 1(b)-(d) the `phase space' parameter
$b \simeq 1$ in all cases while it has to be strongly reduced
assuming the underdamped plasmon ($b \simeq 0.35$).

\section{\label{con}CONCLUSION}

We have demonstrated that the large scale changes found in the
self-energy of the nodal quasiparticles in Bi2201 are in detailed
quantitative agreement with the measured changes in the screening
properties along the c-axis when it is assumed that this
self-energy is mostly due to the scattering of the electrons
against the polar phonons associated with the motions of ions in
the c-direction. This is a direct evidence for the presence and
importance of this type of interaction and in a future publication
we will elaborate further consequences of this unconventional
electron-phonon coupling \cite{smallq:Tom}.

\appendix
\section{EFFECTIVE PLASMON FREQUENCY, $\omega_{p,
\vec{q}}$}

In a coupled layered electron gas, the dielectric function using
the RPA approximation (Eq. (7) in Ref. 9) is given by

\begin{eqnarray}
&&\epsilon(\omega, \vec{q}) = 1 - \frac{S}{\omega (\omega+i 0^+)
}\{ \frac{q^2_{ab}}{q^2} \omega^2_{p, ab}p^0\left( \frac{v_F
q_{ab}}{\omega}, \frac{v_c q_{c}}{\omega} \right) \nonumber \\&& +
\frac{q^2_{c}}{q^2} \omega^2_{p, c}p^0\left( \frac{v_c
q_{c}}{\omega}, \frac{v_F q_{ab}}{\omega} \right) \} \nonumber
\end{eqnarray}
where S is a form factor, $v_c$ is the effective velocity
perpendicular to the layers, and $p^0$ is the function
\begin{equation}
p^0(a,b)\equiv\frac{2}{\pi a} \int^\pi_0 \left\{ (1-a \cos
\phi)^2-b^2\right\}^{-1/2}\cos \phi d\phi \nonumber
\end{equation}

And, hence the effective 3D plasmon frequency $\omega_{p,
\vec{q}}$ may be written as
\begin{eqnarray}
&&\omega^2_{p,\vec{q}} = \frac{q^2_{ab}}{q^2} \omega^2_{p,
ab}p^0\left( \frac{v_F q_{ab}}{\omega}, \frac{v_c q_{c}}{\omega}
\right)  + \frac{q^2_{c}}{q^2} \omega^2_{p, c}p^0\left( \frac{v_c
q_{c}}{\omega}, \frac{v_F q_{ab}}{\omega} \right) \nonumber
\end{eqnarray}
where in the limit $\vec{q}\rightarrow0$, $S\rightarrow1$ and
$p^0\rightarrow1$.

\section{EFFECTIVE INTERACTION, $W_{eff}$}

The detailed derivations of the following can be read from Ref.
13. When considering the single-process scattering of two
electrons: i) by the unscreened electron-electron interaction and
ii) by sending a phonon from one to the other, the combined
interaction $W^0$ can be written as:
\begin{equation}
W^0 (\vec{q}, \omega ) = v^{\infty}_q +
V_{ph}(\vec{q},\omega)\nonumber\label{bareint}
\end{equation}
where $v^{\infty}_q = 4 \pi e^2 / ( q^2 \epsilon_{\infty})$ is a
bare Coulomb part and $V_{ph}$ is a phonon mediated part.

This phonon mediated part (same as Eq.( \ref{bphonprop}) ) is
given by,
\begin{equation}
V_{ph} (\vec{q}, \omega ) = M^2_{q} D^0 (q, \omega)\nonumber
\end{equation}
where $M_q$ is a bare EP vertex and $D^0$ is a bare phonon
propagator.

Next, when considering all the possible of multiple-process
scattering, the effective interaction in the classic RPA form,
$W_{eff}$, can be written as:

\begin{equation}
W_{eff} (\vec{q}, \omega ) = \frac {W^0 (\vec{q}, \omega )} {1 -
W^0 (\vec{q}, \omega ) P_e ( \vec{q}, \omega )} = \frac { v_q } {
\epsilon_{tot} (\vec{q}, \omega)  }\nonumber
\end{equation}
where $\epsilon_{tot} (\vec{q}, \omega)=\epsilon_{\infty} - v_q
P_e ( \vec{q}, \omega ) + \epsilon_{ph} ( \vec{q}, \omega )$, $P_e
( \vec{q}, \omega )$ is an electronic propagator and
$\epsilon_{ph} ( \vec{q}, \omega )$ is a phonon polarizability
given by Eq. (\ref{phoneps}).

To explicitly separate the electron-electron part of the effective
interaction, $W_{eff}$ can be rewritten to be the same as Eq. (
\ref{effint}) or:

\begin{equation}
W_{eff} (\vec{q}, \omega) = \frac { v_q } { \epsilon_{tot}
(\vec{q}, \omega)  } = \frac{ v_q} {\epsilon_{el} ( q, \omega)} +
W_{sc-ph}\nonumber
\end{equation}
where $W_{sc-ph}$ is the screened EP interaction given by Eq.
(\ref{phonint}) and the renormalized phonon propagator is given by
Eq. (\ref{phonprop}).

\section{NUMERICAL FACTOR, $b$}

In the insulating case ($\epsilon_{el} = \epsilon_{\infty}$), from
Eq. (\ref{eliasdef}) and (\ref{elias2eps}), $b$ can be written as

\begin{equation}
b = \frac{\frac { 1} { (2\pi)^3} \int \frac{d^2 k'} {v_F} |
M_{\vec{k}-\vec{k'}} |^2 Im D^0 (\vec{k} -\vec{k'},
\omega)}{\frac{|M^2_{q \simeq 0}|}{v_{q \simeq 0}} Im D^0 (\omega)
} \nonumber
\end{equation}
In the overdoped case, replace $M$ with
$\epsilon_{\infty}M/\epsilon_{el}$ and $D^0$ with $D$ in the above
equation.

When momentum dependences are weak, the factor $b$ can be
approximated by

\begin{equation}
b \approx \frac { v_{q \simeq 0}} { (2\pi)^3} \int \frac{d^2 k'}
{v_F} \nonumber
\end{equation}

\begin{acknowledgments}
We thank D. van der Marel and A. Damascelli for enlightening
discussions. SSRL is operated by the DOE Office of Basic Energy
Science under Contract No. DE-AC02-76SF00515. ARPES measurements
at Stanford were supported by NSF DMR-0304981 and ONR
N00014-04-1-0048. W.M. acknowledges DPST for financial support.
T.P.D. would like to thank ONR N00014-05-1-0127, NSERC, and
Alexander von Humboldt foundation. J.Z. acknowledges the support
of the Fulbright foundation.
\end{acknowledgments}


\begin{thebibliography}{99}

\bibitem{gencaxis}
For the doping-dependent in-plane and out-of-plane resistivity of
Bi2201 samples, respectively, Y. Ando, S. Komiya, K. Segawa, S.
Ono, and Y. Kurita Phys. Rev. Lett. {\bf 93}, 267001 (2004); A. N.
Lavrov, Y. Ando, and S. Ono, Europhys. Lett. {\bf 57}, 267 (2002).

\bibitem{MullerBednorz}
K. A. Muller and J. G. Bednorz, Science \textbf{237}, 1133 (1987)

\bibitem{Falter}
C. Falter, G. A. Hoffmann, and F. Schnetgoke, J. Phys.: Condens.
Matter \textbf{14}, 3239 (2002).

\bibitem{Non:experiment}
W. Meevasana, N. J. C. Ingle, D. H. Lu, J. R. Shi, F. Baumberger,
K. M. Shen, W. S. Lee, T. Cuk, H. Eisaki, T. P. Devereaux, N.
Nagaosa, J. Zaanen, and Z.-X. Shen, Phys. Rev. Lett. \textbf{96},
157003 (2006).

\bibitem{breathing}
For a recent review, see L. Pintschovius, Phys. Stat. Sol.(b)
{\bf 242}, 30 (2005).

\bibitem{Zhou}
X. J. Zhou {\it et al.}, Phys. Rev. Lett. \textbf{95}, 117001
(2005); T. Cuk {\it et al.}, Phys. Stat. Sol. (b) {\bf 242}, 11
(2005). A. Lanzara \emph{et al.}, Nature \textbf{412}, 510 (2001).

\bibitem{charge_ordering}
R. J. McQueeney, Y. Petrov, T. Egami, M. Yethiraj, G. Shirane, and
Y. Endoh, Phys. Rev. Lett. \textbf{82}, 628 (1999).

\bibitem{ishihara}
S. Ishihara and N. Nagaosa, Phys. Rev. B {\bf 69}, 144520 (2004).

\bibitem{PlasmonInOut:Marel}
D. van der Marel and J.H Kim, J. Phys. Chem. Sol. \textbf{56},
1825 (1995)

\bibitem{InplanePlasma:Marel}
A.A. Tsvetkov, J. Schutzmann, J. I. Gorina, G. A. Kaljushnaia, and
D. van der Marel, Phys. Rev. B \textbf{55}, 14152 (1996).

\bibitem{LSCO:Tajima}
S. Uchida, K. Tamasaku and S. Tajima, Phys. Rev. B \textbf{53},
14558 (1996).

\bibitem{Falikov}
H. Morawitz, I. Bozovic, V.Z. Kresin, G. Rietveld, and D. van der Marel, Z. Phys. B {\bf 90},
277 (1993) and ref's therein.

\bibitem{Mahan}
G.D. Mahan, \emph{Many Particle Physics} (Plenum Press, New York,
1981), ch.6, p. 560-566, 569-570 and 586-594.

\bibitem{eliashberg:Grimvall}
G. Grimvall and E. Wohlfarth, \emph{The Electron-Phonon
Interaction in Metals}, (North-Holland, New York, 1981), ch.5, p.
98-107.

\bibitem{smallq:Tom}
T.P. Devereaux \emph{et al.}, to be published.

\bibitem{LossFunction:Marel}
A.A. Tsvetkov, D. Dulic, D. van der Marel, A. Damascelli, G. A.
Kaljushnaia, J. I. Gorina, N. N. Senturina, N. N. Kolesnikov, Z.
F. Ren, J. H. Wang, A. A. Menovsky, and T. T. M. Palstra, Phys.
Rev. B \textbf{60}, 13196 (1999).

\bibitem{Classicplas} M.E. Kim, A. Das and S.D. Senturia,
Phys. Rev. B {\bf 18}, 6890 (1978); B.A. Sanborn, Phys. Rev. B
{\bf 51}, 14256 (1995).

\end{thebibliography}
\end{document}